\documentclass{npw}%[12pt]{iopart}
\usepackage{graphicx}
\usepackage{amsmath}
\usepackage{url}
\usepackage{pstricks}

\begin{document}

\title{Fitting Skyrme functionals using linear response theory}
\author{
  A. Pastore\email{pastore@ipnl.in2p3.fr},
  D. Davesne\email{davesne@ipnl.in2p3.fr},
  K. Bennaceur\email{bennaceur@ipnl.in2p3.fr},
  J. Meyer\email{jmeyer@ipnl.in2p3.fr}  \\
  \it Universit\'e de Lyon, F-69003 Lyon, France; \\
  \it Universit\'e Lyon 1,
      43 Bd. du 11 Novembre 1918, F-69622 Villeurbanne cedex, France\\
  \it CNRS-IN2P3, UMR 5822, Institut de Physique Nucl{\'e}aire de Lyon \\
  V. Hellemans\email{veerle.hellemans@ulb.ac.be}\\
   \it Universit\'e Libre de Bruxelles, Physique Nucl\'eaire Th\'eorique, \\
   \it CP229, BE-1050 Bruxelles, Belgium
   }
\pacs{XXX}
\date{}
\maketitle

%%%%%%%%%%%%%%%%%%%%%%%%%%%%%%%%%%%%%%%%%%%%%%%%%%%%%%%%%%%%%%%%%%%%%%%%%%%%%%

\begin{abstract}

Recently, it has been recently shown that the linear response theory in symmetric
nuclear matter can be used as a tool to detect finite size instabilities
for different Skyrme functionals~\cite{Pastore12b,pastoreLetter}. In particular
it has been shown that there is a correlation between the density
at which instabilities occur in infinite matter and the instabilities
in finite nuclei.

In this article we present a new fitting protocol that uses this correlation
to add new additional constraint  in Symmetric Infinite Nuclear Matter in order to
ensure the stability of finite nuclei against matter fluctuation in all
spin and isospin channels. As an application, we give the parameters set for
a new Skyrme functional which includes central
and spin-orbit parts and which is free from instabilities by construction.

\end{abstract}

%%%%%%%%%%%%%%%%%%%%%%%%%%%%%%%%%%%%%%%%%%%%%%%%%%%%%%%%%%%%%%%%%%%%%%%%%%%%

\section{Introduction}

The formalism of the Linear Response (LR) theory has been presented in a
recent series of articles written by the NESQ
collaboration~\cite{Davesne09,Davesne12,Davesne12b} for  the case of a general
Skyrme functional~\cite{Lesinski07} including spin-orbit and tensor
interactions.
Such a formalism can be adopted as a useful tool to detect instabilities that
plague Skyrme functionals~\cite{Pastore12b,pastoreLetter,Lesinski06}.

In the present article, we present an improved fitting protocol to build Skyrme
functionals, that, for the first time, includes the stability of Symmetric Nuclear Matter (SNM) up to a given density as an additional constraint.
Such method is more powerful than the standard constraints on the Landau
parameters, since the latter are only valid in the so called
\emph{long-wavelength} limit and are therefore not able to detect finite-size
instabilities.

An example of this kind of instabilities has been revealed
by Lesinski \emph{et al.}~\cite{Lesinski06} and was shown to be related
to large values of $C_1^{\Delta\rho}$, the coupling constant of the
term $\rho_{1}\Delta\rho_{1}$ in the Skyrme functional~\cite{Lesinski07a}.
This coupling constant, if not properly constrained during the fitting
procedure, can drive the merit function $\chi^{2}$ used to adjust the
parameters of the interaction in a region of matter fluctuations that give
unphysical properties for the ground state of some nuclei.
Since this gradient term does not contribute to the Landau parameters,
this kind of instabilities is only detectable trough the LR method.

The article is organized as follows: in Section~\ref{edf} we recall the main
properties of a Skyrme functional and in Section~\ref{lr:theory} we briefly
remind the LR formalism; in Section~\ref{results} we present the results of
our new fitting method and finally we give our conclusions in
Section~\ref{concl}.

\section{The energy functional}\label{edf}

The density-dependent Skyrme interaction considered in the present article
has the standard form
\begin{align}
v(\mathbf{R},\mathbf{r})
&= t_{0}(1+x_{0}\hat{P}_{\sigma})\delta(\mathbf{r})\nonumber\\
&+ \tfrac{1}{6}\,t_{3}(1+x_{3}\hat{P}_{\sigma})
\rho_{0}^{\alpha}(\mathbf{R})\delta(\mathbf{r})\nonumber\\
&+ \tfrac{1}{2}t_{1}(1+x_{1}\hat{P}_{\sigma})
\left[ {\mathbf{k}\smash{'}}^2\delta(\mathbf{r})
+\delta(\mathbf{r})\mathbf{k}^{2}\right]\nonumber\\
&+ t_{2}(1+x_{2}\hat{P}_{\sigma}) \mathbf{k}'\cdot
 \delta(\mathbf{r})\mathbf{k}\nonumber\\
&+ \mathrm{i}\,W_{0}(\hat{\boldsymbol\sigma}_{1}
+\hat{\boldsymbol\sigma}_{2})
\cdot \left[ \mathbf{k}'\times \delta(\mathbf{r})\mathbf{k}\right]\,,
\end{align}
where we used $\mathbf{r}\equiv \mathbf{r}_{1}-\mathbf{r}_{2}$ and
$\mathbf{R}=\frac{1}{2}\left(\mathbf{r}_{1}+\mathbf{r}_{2}\right)$ for
the relative and center of mass coordinates, respectively.
We also defined $\hat{P}_{\sigma}$ the spin exchange operator, while
$\mathbf{k}=-\frac{\mathrm i}{2}(\nabla_{1}-\nabla_{2})$ is the relative momentum
acting on the right and $\mathbf{k}'$ its complex conjugate acting on the left.
Finally $\rho_{0}(\mathbf{R})$ is the scalar-isoscalar density of the system.
The Energy Density Functional (EDF) is derived by keeping all the terms
except the Coulomb exchange one which is treated using the Slater
approximation. The center of mass correction is approximated by its one-body
part.

\section{Linear response theory}\label{lr:theory}

\begin{figure*}
\begin{center}
\includegraphics[angle=-90,width=0.45\textwidth]{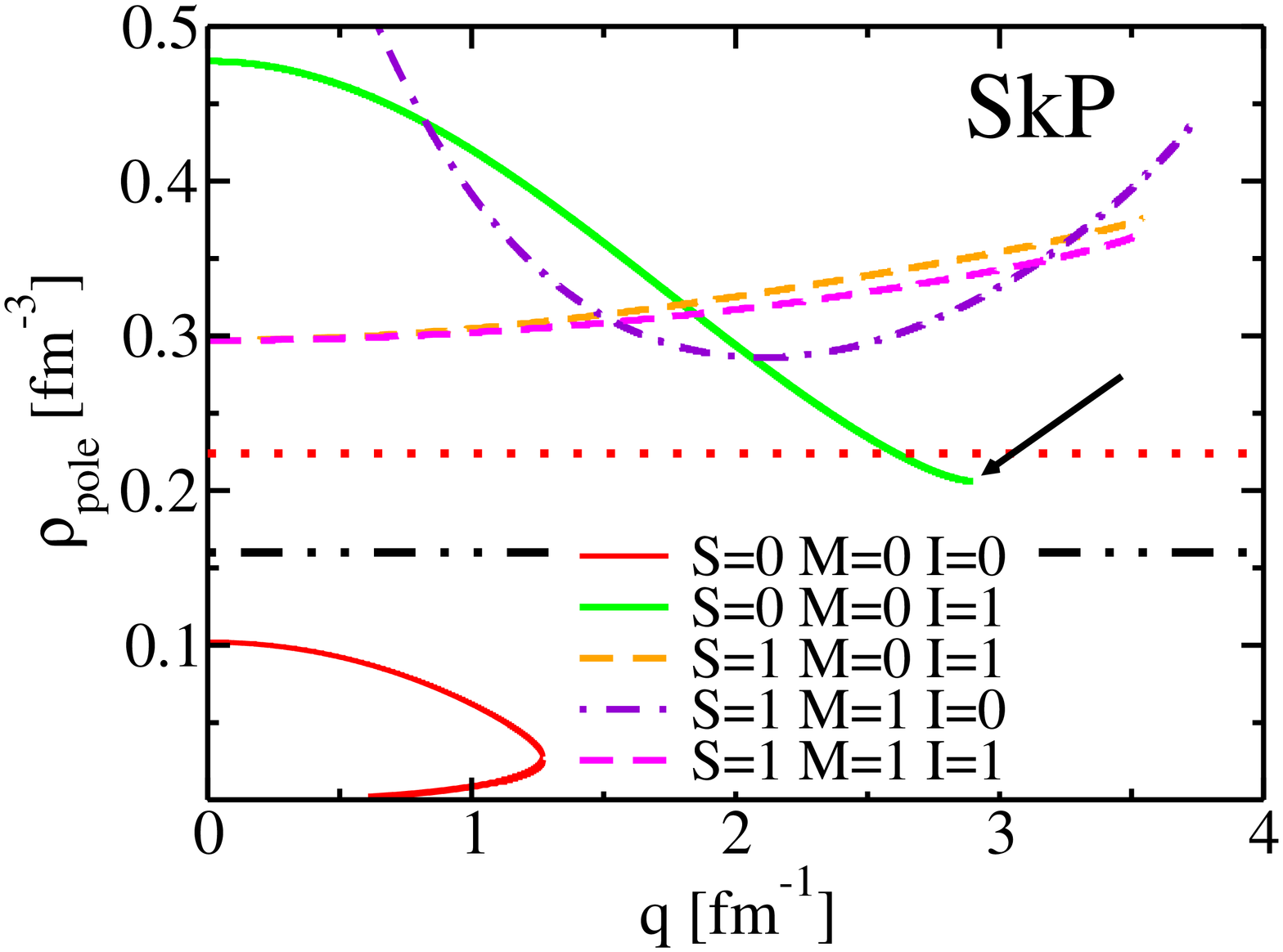}
\includegraphics[angle=-90,width=0.45\textwidth]{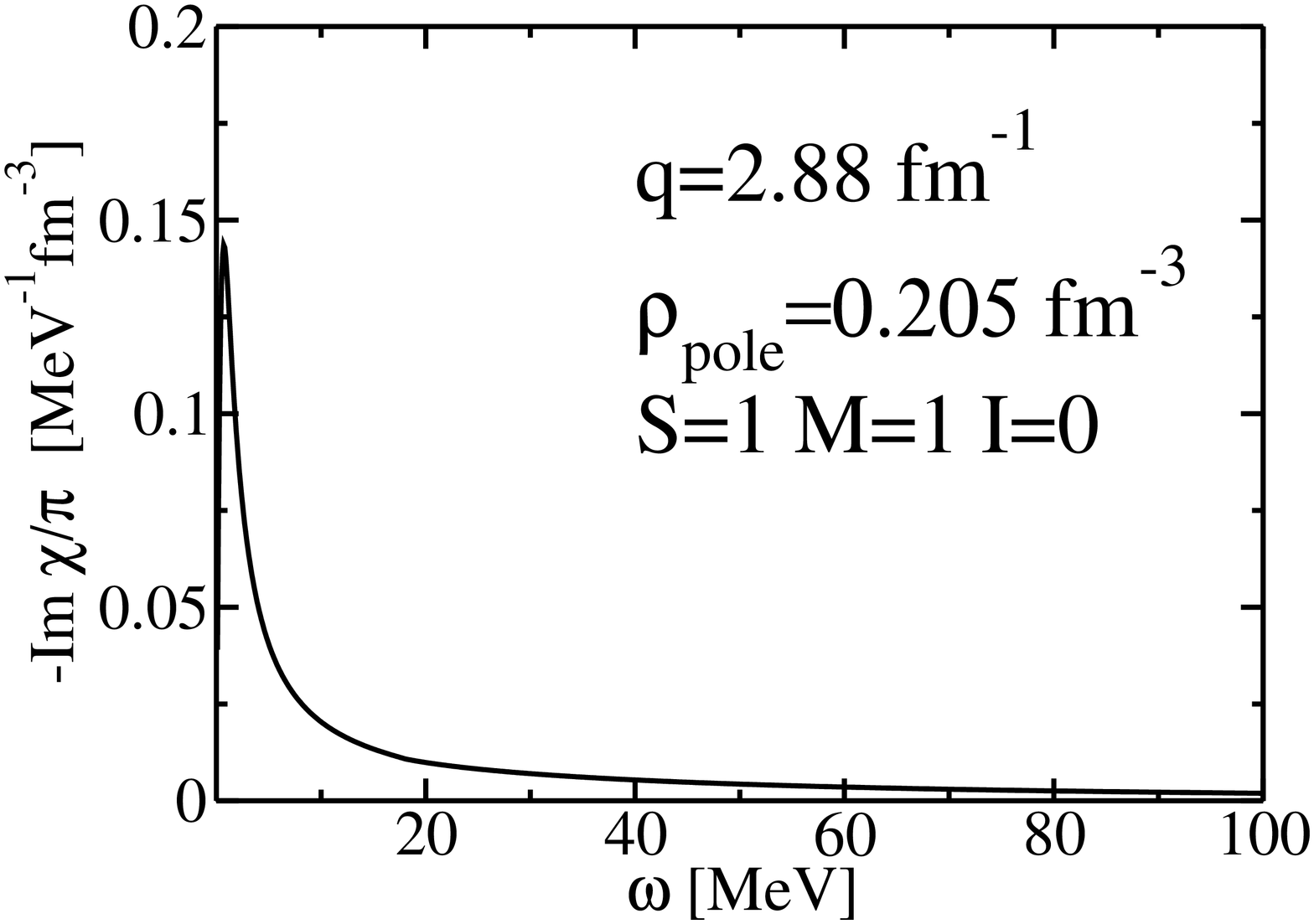}
\end{center}
\caption{(Colors online). In the left panel, we show the position of the poles
in SNM for the SKP functionals. The horizontal black dotted-dashed line represents
the saturation density of the system, $\rho_\mathrm{sat}$, while the red dotted
line is placed at $1.4\,\rho_\mathrm{sat}$. In the right panel, we show the response
function in SNM for the value of density and transferred momentum indicated
by the black arrow in left panel. See the text for details.}
\label{critical:SKP}
\end{figure*}

The Random Phase Approximation (RPA) response function in each channel
$(\alpha)=(S,M,I)\equiv$ (spin, projection of the spin and isospin) is
obtained by solving the Bethe-Salpeter equations for the correlated Green
functions $G_\mathrm{RPA}^{(\alpha)}$ written as
\begin{eqnarray}\label{Bethesalpeter}
&&G_\mathrm{RPA}^{\alpha}(q,\omega,\mathbf{k}_1)=G_\mathrm{HF}(q,\omega,\mathbf{k}_1)+G_\mathrm{HF}(q,\omega,\mathbf{k}_1)\nonumber\\
&&\times\sum_{\alpha'}\int\!\frac{\mathrm d^3 k_2}{(2\pi)^3}\,
V_{ph}^{\alpha;\alpha'}(q,\mathbf{k}_1,\mathbf{k}_2)\,
G^{\alpha'}_\mathrm{RPA}(q,\omega,\mathbf{k}_2),
\end{eqnarray}
where $G_\mathrm{HF}(q,\omega,\mathbf{k}_1)$ is the unperturbed \emph{particle-hole} propagator, and $V_{ph}^{\alpha;\alpha'}(q,\mathbf{k}_1,\mathbf{k}_2)$ is the residual interaction obtained through second functional derivative techniques.
We refer to articles ~\cite{GarciaRecio,Davesne09,Davesne12} for more details.
Once Eqs.~(\ref{Bethesalpeter}) are solved, we can calculate the response
function of the system with
\begin{equation}
\chi^{\alpha}(\omega,q)
=4\int\!\frac{\mathrm d^3k}{(2\pi)^{3}}\,
G_\mathrm{RPA}^{\alpha}(q,\omega,\mathbf{k}),
\end{equation}
where $\omega$ is the transferred energy and $q$ is the transferred momentum. 
As usual we chose the momentum $q$ oriented along the $z-$axis of the reference frame.
We also use natural units $\hbar=c=1$ to simplify the notations.
When an instability appears in the infinite system, it manifests itself by
a pole in the response function $\chi^{\alpha}(\omega,q)$ at zero energy.

In Figure~\ref{critical:SKP} we show the positions of the poles in SNM for
 the Skyrme SkP~\cite{doba84} functional. 
We observe that in the channel {$(S,M,I)=(0,0,1)$},
there is a pole for $\rho_\mathrm{pole}=0.205$~fm$^{-3}$ and $q=2.88$~fm$^{-1}$.
Lesinski {\em et al.}~\cite{Lesinski06} showed that such functional is not
stable: performing high precision Hartree-Fock calculations (HF) using a
sufficient number of iterations, the code converges to a solution where
protons and neutrons are separated in clusters.
The same conclusion was found for  other Skyrme functionals, for example the
LNS~\cite{giai:lns} one.
Such unphysical configurations were discovered to be related to an excessive
contribution of the term proportional to $\rho_{1}\Delta\rho_{1}$ in the
functional.

Performing extensive HF calculations for even-even~\cite{pastoreLetter}
and odd-even nuclei~\cite{schunck}, we have found that there exists a
relation  between the position of the poles of the response function in SNM
and the instabilities in finite nuclei.
To guarantee that a nucleus will not manifest finite-size instabilities we have
imposed the  empirical constraint
\begin{equation}\label{crit:stability}
\rho_\mathrm{pole}>1.4\,\rho_\mathrm{sat},
\end{equation}
where $\rho_\mathrm{sat}$ is the saturation density of SNM.
Such relation has to be satisfied for all the spin/isospin channels.
The only instability that we allow to exist in the density region
$\rho<1.4\,\rho_\mathrm{sat}$ is the spinodal instability, that manifests
in the $(0,0,0)$ channel at low densities and low transferred momenta and
characterizes the liquid-gas phase transition of nuclear matter.
Using such criterion allows to quickly test large sets of Skyrme functionals.
This is a great advantage compared to self consistent cranking calculations~\cite{veerle12} and systematic quasi-particle blocking calculations~\cite{schunck10} used to reveal the
instabilities in spin channels.

Usually, the time-odd terms of the functional are not directly constrained to observables. Indirect constraints may come from the fact that
some functionals have been constructed with the
requirements that the time odd terms lead to Landau
parameters satisfying the conditions
\begin{equation}
\frac{F^{}_{\ell}}{2\ell+1}>1\,,~~~~\frac{G^{}_{\ell}}{2\ell+1}>1,
\end{equation}
for $\ell=0,\,1$
and equivalent ones for $F'_{\ell}$ and $G^{'}_{\ell}$. These
conditions are used to prevent the appearance of global spin and/or isospin
instabilities, but do not prevent the formation
of finite size instabilities due to the gradient terms such as
$\mathbf{s}_{t}\Delta \mathbf{s}_{t}$ and 
$(\nabla \mathbf{s}_{t})^{2}$ for isospin $t=0,1$.

\begin{figure*}
\begin{center}
\includegraphics[angle=-90,width=0.45\textwidth]{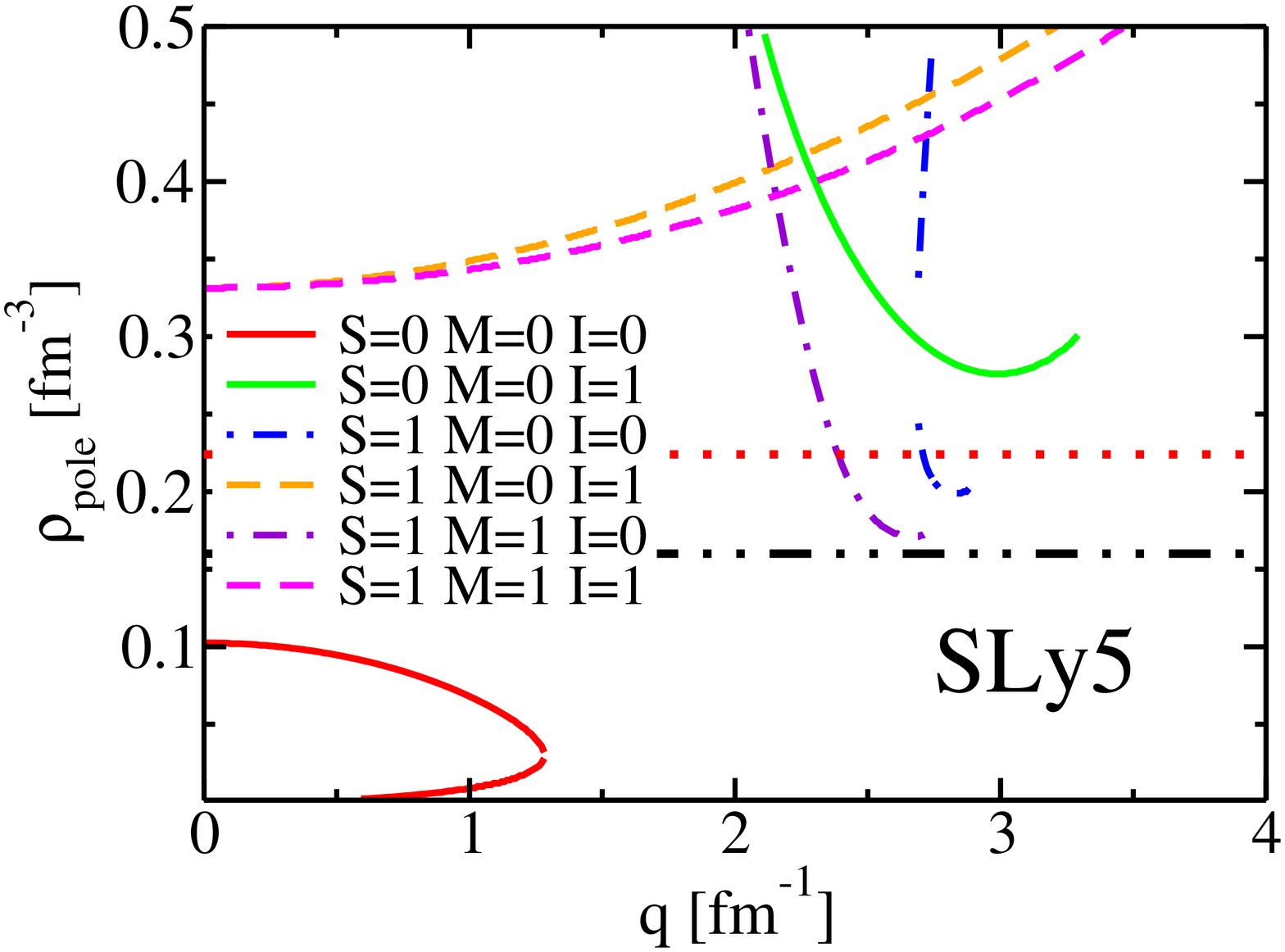}
\includegraphics[angle=-90,width=0.45\textwidth]{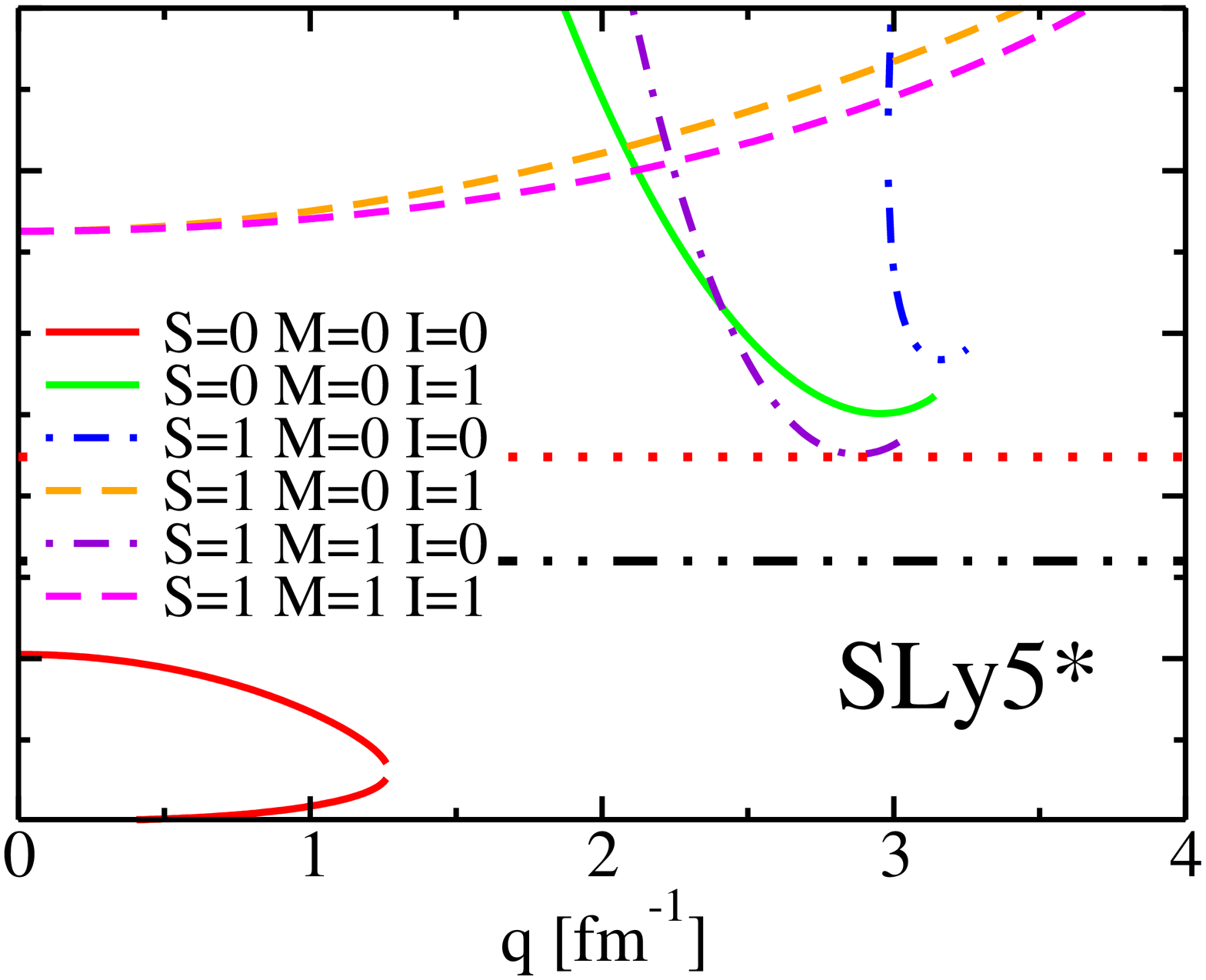}\\
\end{center}
\caption{(Colors online). Position of the poles in SNM for two different functionals. The horizontal black dotted-dashed line represents the saturation density of the system, $\rho_{sat}$, 
while the red dotted line indicates $1.4\rho_{sat}$ and it represents the constraint given in Eq.\ref{crit:stability}. See text for details. }
\label{critical}
\end{figure*}

An example of such spin instability has been found with the
SLy5~\cite{Chabanat98} functional. In Figure~\ref{critical} we show the
position of the poles of the response function in SNM. We notice that there
is a pole in the channel $(1,1,0)$ close to saturation density that does not
respect the criterium given in Eq.~(\ref{crit:stability}). Performing
Time-Dependent-Hartree-Fock (TDHF) calculations, Fracasso
\emph{et al.}~\cite{fracasso12} have shown that the functional SLy5 is 
indeed unstable against spin-dipole excitations.

\section{Results}\label{results}

Using the results discussed in the previous section,
we now present a new fitting protocol to build stable Skyrme functionals.
We use a modern version of the Saclay-Lyon fitting protocol similar to
the one presented in the recent article of Washiyama {\em et al.}~\cite{kowei}.
On top of such protocol we add the constraint given by
Eq~(\ref{crit:stability}).
This constraint is enforced at each iteration of the fit and does not
significantly change the execution time since, as already discussed in
ref.~\cite{Davesne12}, since the expressions of the response functions are
analytical.

Using the SLy5 functional as a starting point,
we decided to build a new functional with
similar properties, but without instabilities in the spin channel.
The resulting functional has been called SLy5* and its coupling constants
are reported in Table~\ref{tab:forces}. In the same table we compare
nuclear matter properties at the saturation point for the new SLy5* 
and the original SLy5, showing that they all are essentially equals
but the Thomas-Reiche-Kuhn enhancement factor $\kappa_v$
which is larger with SLy5*. The value obtained with SLy5* is 
in agreement with the empirical value~\cite{Meyer2003} close to $\kappa_v^{emp}\approx0.4$.

\begin{table}[htbp]
\begin{center}
\begin{tabular}{l|r@{.}lr@{.}l}
\hline
\hline
\multicolumn{1}{c|}{Parameter}  & \multicolumn{2}{c}{SLy5*} &  \multicolumn{2}{c}{SLy5}\\
\hline  
 $t_{0}$ [MeV\,fm$^{3}$] & -2495&310 & -2484&880\\
 $t_{1}$ [MeV\,fm$^{5}$] & 484&020 &483&130\\
 $t_{2}$ [MeV\,fm$^{5}$] & -469&480 &-549&400\\
 $t_{3}$ [MeV\,fm$^{3+3\alpha}$]&13867&430 &13763&000\\
 $x_{0}$&0&620 &0&778\\
 $x_{1}$ &-0&086 &-0&328\\
 $x_{2}$ &-0&947 &-1&000\\
 $x_{3}$ &0&934 &1&267\\
 $\alpha$ & \multicolumn{2}{r}{1/6} & \multicolumn{2}{r}{1/6}\\
 W [MeV\,fm$^{5}$]& 120&250 & 126&000\\
% $J^{2}$ & yes &  yes\\
\hline
\hline
$E/A$ [MeV] &\multicolumn{2}{r}{-16.02} &\multicolumn{2}{r}{-15.98} \\
$\rho_\mathrm{sat}$ [fm$^{-3}$] & \multicolumn{2}{r}{0.160}& \multicolumn{2}{r}{0.160}\\
$m^{*}/m$ & \multicolumn{2}{r}{0.700} & \multicolumn{2}{r}{0.697} \\
$K_{\infty}$ [MeV] & \multicolumn{2}{r}{229.8} & \multicolumn{2}{r}{229.9}\\
$a_I$~[MeV] & \multicolumn{2}{r}{32.01} & \multicolumn{2}{r}{32.03} \\
$\kappa_v$  & \multicolumn{2}{r}{0.418} & \multicolumn{2}{r}{0.250} \\
\hline
\hline
\end{tabular}
\caption{Parameters of the Skyrme functionals used in the article. In the last
lines we give the main SNM properties around saturation for these functionals:
binding energy per nucleon $E/A$, saturation density $\rho_\mathrm{sat}$,
isoscalar effective mass $m^*/m$, incompressibility $K_\infty$, symmetry
energy coefficient $a_I$ and the Thomas-Reiche-Kuhn enhancement factor
$\kappa_v$.}
\label{tab:forces}
\end{center}
\end{table}

\subsection{Equations of State}

In order to compare the isoscalar and isovector properties of
the functionals SLy5 and SLy5* we have calculated the equation of state (EoS)
of SNM and PNM as well as the neutron effective mass in these to
extreme media.

\begin{figure}[bhtp]
\begin{center}
\includegraphics[angle=-90,width=\columnwidth]{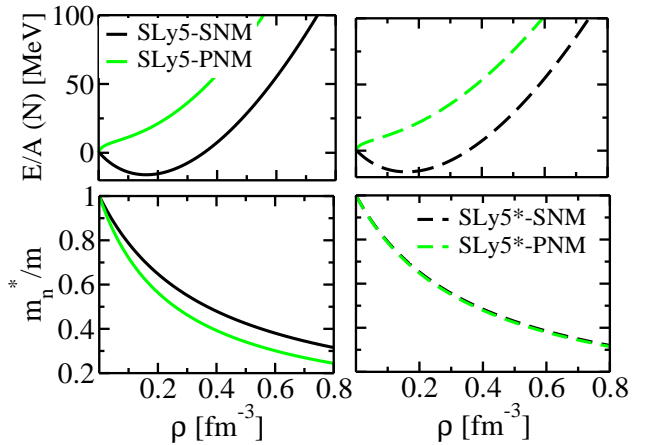}
\end{center}
\caption{(Colors online) In the upper panels, we show the equation of state
in SNM and PNM for the two forces used in the article:  SLy5 (left panel) and  SLy5* (right panel). In the lower panels,
the neutron effective mass is depicted for the two systems. See text for
details.}
\label{eos}
\end{figure}

In the upper panel of Figure~\ref{eos}, we show the equations of state
for SNM and Pure Neutron Matter (PNM). We observe a very good agreement
between the results given by two functionals. In SNM the 
two curves stay on top of each other and in PNM this is the case 
up to 2 to 3 times saturation density.

In the lower panel of the same figure, we report the evolution of the neutron
effective mass in SNM and PNM.
In SNM,  the curves that give the neutron effective mass
$m^{*}/m$ as a function of the density stay on top of each and can
not be distinguished.
In the PNM case, the neutron effective mass obtained from SLy5 is
systematically lower than the effective mass in SNM.

One can also observe that SLy5* almost gives the same evolution
for the neutron effective mass in SNM and in PNM.
The origin of this property can be clarified by writing the expressions
for the neutron effective masse in the two media in terms of coupling
constants~\cite{Lesinski06}
\begin{eqnarray}
\label{eff:mass}
\frac{\hbar^{2}}{2m^{*}_{n}}&=&\frac{\hbar^{2}}{2m}+C^{\tau}_{0}\rho\phantom{++C^{\tau}_{1}} \;\;\;\;\; \text{(SNM),}\\
\frac{\hbar^{2}}{2m^{*}_{n}}&=&\frac{\hbar^{2}}{2m}+(C^{\tau}_{0}+C^{\tau}_{1})\rho\;\;\;\;\; \text{(PNM).}
\end{eqnarray}
These coupling constants are $C^{\tau}_{0}=55.18$~MeV\,fm$^{5}$ and
$C^{\tau}_{1}=1.23$~MeV\,fm$^{5}$ for the new SLy5* and
$C^{\tau}_{0}=56.25$~MeV\,fm$^{5}$ and $C^{\tau}_{1}=23.95$~MeV\,fm$^{5}$
for the original SLy5.
We observe that while the two isoscalar coupling constants are very similar
the isovector one are very different. This, according to Eq.~(\ref{eff:mass}),
explains the difference between the effective masses predicted by
SLy5 and SLy5* as seen in Figure~\ref{eos}. The fact that SLy5* gives
similar effective mass for the neutron in SNM and PNM is due to the
very small value of $C^{\tau}_{1}$ obtained with this interaction.

\subsection{Binding energies}

To further test the quality of the new SLy5* interaction,
we calculated the binding energies of a set of semi-magic nuclei
using the Hartree-Fock-Bogoliubov (HFB) method.
For the {\em particle-hole} channel we use the two functionals SLy5 and SLy5*
while for the {\em particle-particle} channel we used a contact pairing
interaction surface-type.

With such choice we obtain an average pairing gap~\cite{stoitsov} $\Delta_\mathrm{aver}=1.201$~MeV for SLy5 and $\Delta_\mathrm{aver}=1.164$~MeV for SLy5*.
We compare the resulting binding energies with the experimental ones given
by Audi {\em et al.}~\cite{audi}.
We observe that there is no significant qualitative difference between
the two functionals for this set of nuclei. The binding energies predicted
with SLy5* are in average slightly smaller than with SLy5 but
both functional compare with the experimental data with the same
average deviation.

\begin{figure}[htbp]
\begin{center}
\includegraphics[angle=-90,width=\columnwidth]{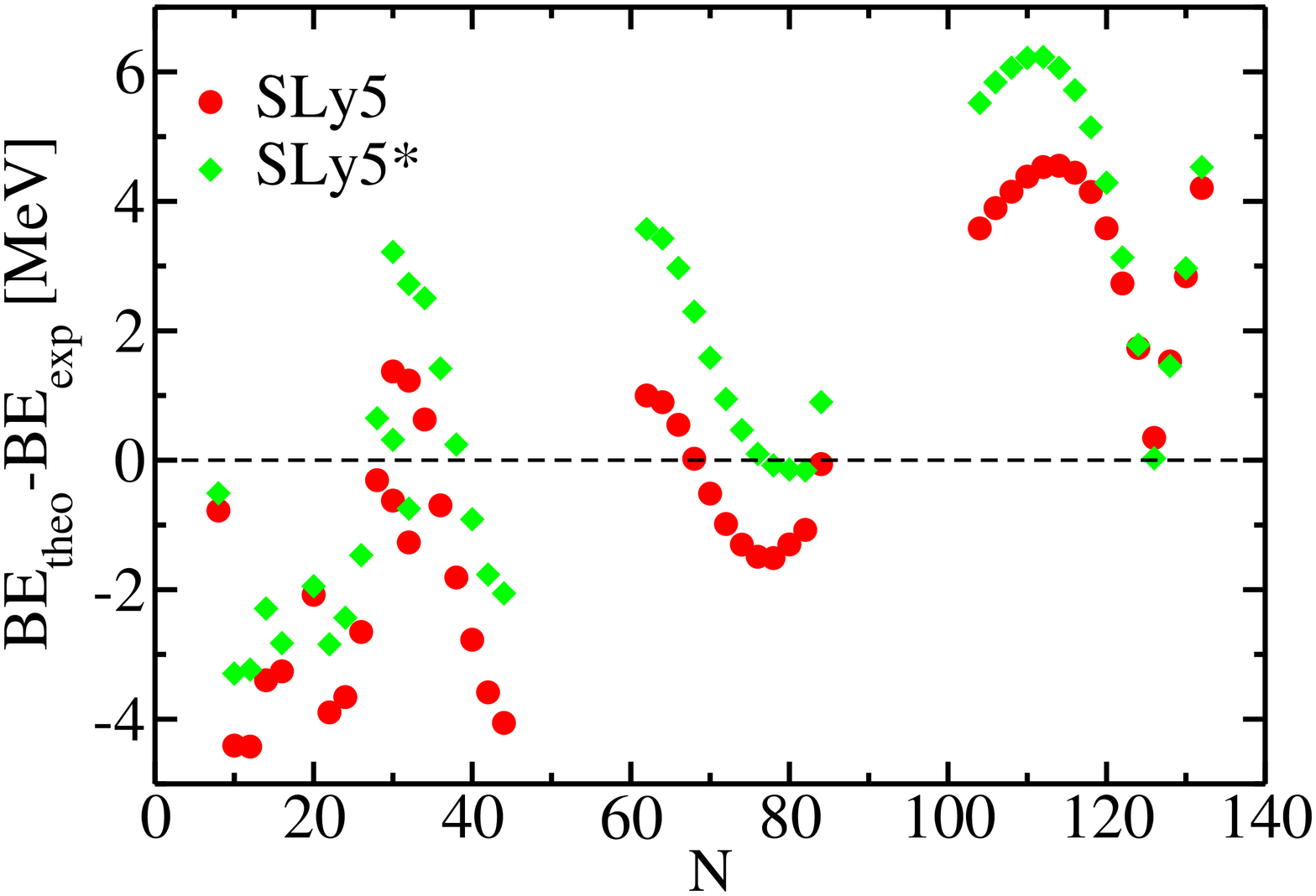}\\
\includegraphics[angle=-90,width=\columnwidth]{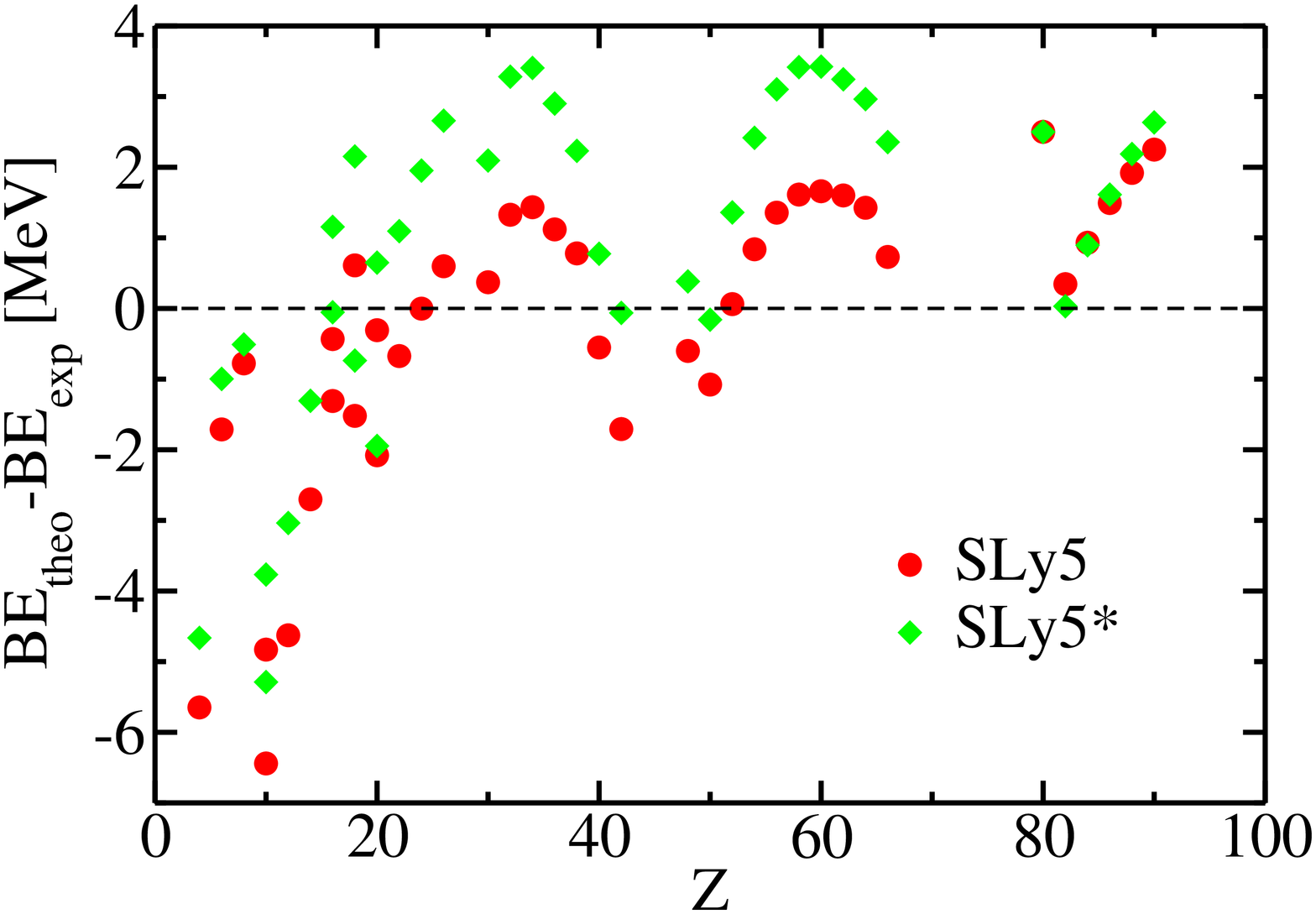}\\
\end{center}
\caption{(Colors online)In the upper panel, we show the difference in biding energies for different isotopic chains of semi-magic nuclei obtained with SLy5 and SLy5*. In the lower panel the same, but for isotonic chains.}
\label{be}
\end{figure}

\subsection{Charge radii}

\begin{figure}[htbp]
\begin{center}
\includegraphics[angle=-90,width=\columnwidth]{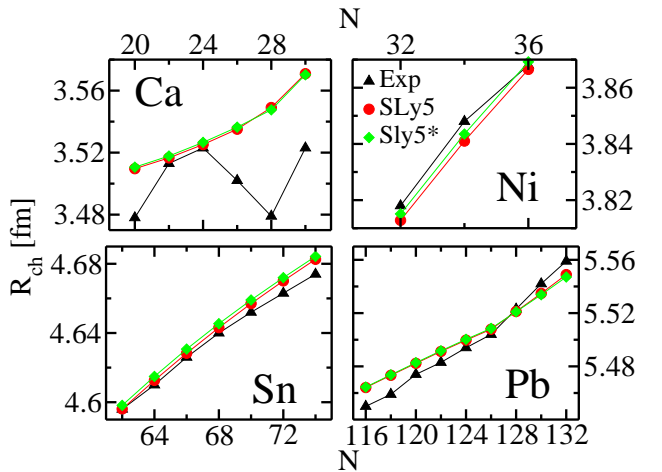}\\
\end{center}
\caption{(Colors online)Comparing charge radii for two different force for four different isotopic chains.}
\label{cradii}
\end{figure}

In Fig.\ref{cradii} we compare the charge radii obtained with the two functionals against the experimental results taken from the UNEDF web
page~\footnote{orph02.phy.ornl.gov/workshops/lacm08/UNEDF/DataSet04.dat\!\!}.
Once again, we observe that SLy5 and SLy5* show the same behavior and the agreement between theory and experiment is of the same level.

\subsection{Stability against spin polarization}

Following the work done in Ref.\cite{veerle12}, we tested the stability of the new force SLy5* in self-consistent cranking calculations.
For this purpose, we calculated a high-spin state of the ground super-deformed band of of $^{194}$Hg,  that is at $J_{z}=54$ (natural units), going up to several thousands iterations.
We observe that SLy5* is perfectly stable and does not display the problems detected with SLy5, which typically leads to systematic non-convergence within few iterations.

\section{Conclusions}\label{concl}

We have shown that, with the inclusion of the LR method into the fitting protocol,
it is possible to build functionals that do not present instabilities.
As a first example, we used this improved protocol to cure the spin instability present in SLy5.
The resulting functional has a comparable level of accuracy as the
original one for the prediction of relevant observables as binding energies
and charge radii for the set of semi-magic nuclei considered here, moreover
most of the relevant infinite nuclear matter properties are unchanged only
the Thomas-Reiche-Kuhn enhancement factor is significantly larger.

Such method clearly demonstrates that it is possible to use the LR
method for additional constraints to build a functional with no finite
size instabilities in infinite nuclear matter and nuclei.

\begin{ack}
This work was supported by the ANR NESQ project (ANR-BLANC 0407). V.H. acknowledges financial support from the F.R.S.-FNRS (Belgium).
The authors also thank T. Duguet, P.-H. Heneen, M. Bender for inspiring and interesting discussions. 
\end{ack}

%\section*{References}


\begin{thebibliography}{99}
\bibitem{Pastore12b} A. Pastore, K. Bennaceur, D. Davesne and J. Meyer, Journal of Mod. Phys. E  5, 1250041, (2012). 
\bibitem{pastoreLetter}A. Pastore, T. Duguet,  K. Bennaceur, D. Davesne, J. Meyer, V. Hellemans, M. Bender  and P.-H. Heenen \emph{in preparation}.
\bibitem{Davesne09}D. Davesne, M. Martini, K. Bennaceur, J. Meyer, Phys. Rev. C \textbf{80}, 024314 (2009); Phys. Rev. C Erratum \textbf{84}, 059904 (2011).
\bibitem{Davesne12}A. Pastore,  D. Davesne,  Y. Lallouet, M. Martini, K. Bennaceur and J. Meyer, Phys. Rev. C \textbf{85}, 054317 (2012).
\bibitem{Davesne12b}A. Pastore,  M. Martini, V. Buridon, D. Davesne,  K. Bennaceur and J. Meyer, Phys. Rev. C \textbf{86}, 044308 (2012).
\bibitem{Lesinski07}T. Lesinski, M. Bender, K. Bennaceur, T. Duguet, and J. Meyer, Phys. Rev. C \textbf{76}, 014312, (2007).
\bibitem{GarciaRecio}C.  Garcia-Recio, J. Navarro, N. Van Giai and L.L. Salcedo,
Ann. Phys. (N.-Y.) \textbf{214}, 293-340 (1992). 
\bibitem{Lesinski06}T. Lesinski, K. Bennaceur, T. Duguet, and J. Meyer, Phys. Rev. C \textbf{74}, 044315 (2006).
\bibitem{Lesinski07a}T. Lesinski, M. Bender, K. Bennaceur, T. Duguet, and J. Meyer, Phys. Rev. C 76, 014312 (2007).
\bibitem{doba84}J. Dobaczewski, H. Flocard and J. Treiner, Nucl. Phys. \textbf{A422} 103-139 (1984).
\bibitem{giai:lns}L.G. Cao, U. Lombardo, C.W. Shen, and Nguyen Van Giai, Phys. Rev. C \textbf{73}, 014313 (2006).
\bibitem{veerle12}V. Hellemans, P.-H. Heenen, and M. Bender, Phys. Rev. C \textbf{85}, 014326 (2012).
\bibitem{schunck}N. Schunck, T. Duguet, T. Lesinski, K. Bennaceur, A. Pastore, and D. Davesne, \emph{in preparation} (2012).
\bibitem{schunck10} N. Schunck, J. Dobaczewski, J. McDonnel, J. Mor\'e, W. Nazarewicz, J. Sarich and M.V. Stoitsov, Phys. Rev. C \textbf{81}, 024316 (2010)
\bibitem{Chabanat98}E. Chabanat, P. Bonche, P. Haensel, J. Meyer and R. Schaeffer, Nucl Phys \textbf{A627}, 710-746  (1997).
\bibitem{fracasso12}S. Fracasso, E.B. Suckling, and P.D. Stevenson, Phys. Rev. C \textbf{86}, 044303, (2012).
\bibitem{kowei}K. Washiyama, K. Bennaceur, B. Avez, M. Bender, P.-H. Heenen, V. Hellemans, arXiv:1209.5258.

\bibitem{stoitsov}
J. Dobaczewski, M.V. Stoitsov, W. Nazarewicz,
AIP Conference Proceedings Volume 726, ed. R. Bijker, R.F. Casten,
A. Frank (American Institute of Physics, New York, 2004) p. 51.

\bibitem{audi}
G. Audi, A.H. Wapstra and C. Thibault, Nucl. Phys. \textbf{A729}, 337 (2003).

\bibitem{Meyer2003}
J. Meyer, Ann. Phys. Fr. 28, 3 (2003) 1-113 
\end{thebibliography}
\end{document}